\RequirePackage{lineno}

\documentclass[preprintnumbers,aps,prd,twocolumn,showpacs,superscriptaddress,groupedaddress,floatfix]{revtex4}

\usepackage[normalem]{ulem}
\usepackage{graphicx}\usepackage{dcolumn}\usepackage{bm}\usepackage{xspace}
\usepackage{multirow}
\usepackage{color}

\usepackage{amssymb,amsmath,mathrsfs}

\usepackage{dcolumn}\newcolumntype{.}{D{.}{.}{-1}}
\newcolumntype{,}{D{,}{\relax}{-1}}
\newcolumntype{+}{D{+}{\,\pm\,}{-1}}
\newcolumntype{^}{D{^}{^}{-1}}

\usepackage{hyperref}

\newcommand{\mt}       {\ensuremath{m_{t}}\xspace}

\newcommand{\ttbar}     {\mbox{$t\bar{t}$}}
\newcommand{\GeV}{\ensuremath{\mathrm{Ge\kern-0.1em V}}\xspace}
\newcommand{\tev}{\ensuremath{\mathrm{Te\kern-0.1em V}}\xspace}
\newcommand{\pt}       {\ensuremath{p_{T}}}
\newcommand{\met}       {\mbox{$\not\!\!p_T$}}

\newcommand{\gev}{\GeV}
\newcommand{\gevcs}{\GeV}
\newcommand{\invfb}{\ensuremath{\mathrm{fb^{-1}}}}

\newcommand{\ppbar}{\ensuremath{p\bar p}}
\newcommand{\sqrts}{\ensuremath{\sqrt s}}
\newcommand{\dil}        {\ensuremath{\ell \ell^{\prime}}\xspace}
\newcommand{\lplus}      {\ensuremath{\ell+{\rm jets}}\xspace}

\newcommand{\RunII}{Run~II}
\newcommand{\RunI}{Run~I}

\def \ee{\ensuremath{ee}}
\def \emu{\ensuremath{e\mu}}
\def \mumu{\ensuremath{\mu\mu}}
\newcommand{\ie}{{i.e.}}
\def\eg{{e.g.}}

\newcommand{\dzero}{D0\xspace}

\makeatletter{}\newcommand{\central}{174.95}
\newcommand{\stat}{0.40}
\newcommand{\syst}{0.64}
\newcommand{\tot}{0.75}

\newcommand{\result} {\ensuremath{\central \pm \stat\,{\rm(stat)} \pm \syst\,{\rm(syst)}~\gevcs }\xspace} 
\newcommand{\resulttot} {\ensuremath{\central \pm \tot{}~\gevcs }\xspace}

\begin{document}

\widetext

\noindent \mbox{FERMILAB-PUB-17-079-E}  \hspace{15mm}\hfill\mbox{{\em Published in Phys.\ Rev.\ D.\ as
DOI:10.1103/PhysRevD.95.112004}}

\title{Combination of  \dzero measurements of the top quark mass }

\makeatletter{}\affiliation{LAFEX, Centro Brasileiro de Pesquisas F\'{i}sicas, Rio de Janeiro, RJ 22290, Brazil}
\affiliation{Universidade do Estado do Rio de Janeiro, Rio de Janeiro, RJ 20550, Brazil}
\affiliation{Universidade Federal do ABC, Santo Andr\'e, SP 09210, Brazil}
\affiliation{University of Science and Technology of China, Hefei 230026, People's Republic of China}
\affiliation{Universidad de los Andes, Bogot\'a, 111711, Colombia}
\affiliation{Charles University, Faculty of Mathematics and Physics, Center for Particle Physics, 116 36 Prague 1, Czech Republic}
\affiliation{Czech Technical University in Prague, 116 36 Prague 6, Czech Republic}
\affiliation{Institute of Physics, Academy of Sciences of the Czech Republic, 182 21 Prague, Czech Republic}
\affiliation{Universidad San Francisco de Quito, Quito 170157, Ecuador}
\affiliation{LPC, Universit\'e Blaise Pascal, CNRS/IN2P3, Clermont, F-63178 Aubi\`ere Cedex, France}
\affiliation{LPSC, Universit\'e Joseph Fourier Grenoble 1, CNRS/IN2P3, Institut National Polytechnique de Grenoble, F-38026 Grenoble Cedex, France}
\affiliation{CPPM, Aix-Marseille Universit\'e, CNRS/IN2P3, F-13288 Marseille Cedex 09, France}
\affiliation{LAL, Univ. Paris-Sud, CNRS/IN2P3, Universit\'e Paris-Saclay, F-91898 Orsay Cedex, France}
\affiliation{LPNHE, Universit\'es Paris VI and VII, CNRS/IN2P3, F-75005 Paris, France}
\affiliation{CEA Saclay, Irfu, SPP, F-91191 Gif-Sur-Yvette Cedex, France}
\affiliation{IPHC, Universit\'e de Strasbourg, CNRS/IN2P3, F-67037 Strasbourg, France}
\affiliation{IPNL, Universit\'e Lyon 1, CNRS/IN2P3, F-69622 Villeurbanne Cedex, France and Universit\'e de Lyon, F-69361 Lyon CEDEX 07, France}
\affiliation{III. Physikalisches Institut A, RWTH Aachen University, 52056 Aachen, Germany}
\affiliation{Physikalisches Institut, Universit\"at Freiburg, 79085 Freiburg, Germany}
\affiliation{II. Physikalisches Institut, Georg-August-Universit\"at G\"ottingen, 37073 G\"ottingen, Germany}
\affiliation{Institut f\"ur Physik, Universit\"at Mainz, 55099 Mainz, Germany}
\affiliation{Ludwig-Maximilians-Universit\"at M\"unchen, 80539 M\"unchen, Germany}
\affiliation{Panjab University, Chandigarh 160014, India}
\affiliation{Delhi University, Delhi-110 007, India}
\affiliation{Tata Institute of Fundamental Research, Mumbai-400 005, India}
\affiliation{University College Dublin, Dublin 4, Ireland}
\affiliation{Korea Detector Laboratory, Korea University, Seoul, 02841, Korea}
\affiliation{CINVESTAV, Mexico City 07360, Mexico}
\affiliation{Nikhef, Science Park, 1098 XG Amsterdam, the Netherlands}
\affiliation{Radboud University Nijmegen, 6525 AJ Nijmegen, the Netherlands}
\affiliation{Joint Institute for Nuclear Research, Dubna 141980, Russia}
\affiliation{Institute for Theoretical and Experimental Physics, Moscow 117259, Russia}
\affiliation{Moscow State University, Moscow 119991, Russia}
\affiliation{Institute for High Energy Physics, Protvino, Moscow region 142281, Russia}
\affiliation{Petersburg Nuclear Physics Institute, St. Petersburg 188300, Russia}
\affiliation{Instituci\'{o} Catalana de Recerca i Estudis Avan\c{c}ats (ICREA) and Institut de F\'{i}sica d'Altes Energies (IFAE), 08193 Bellaterra (Barcelona), Spain}
\affiliation{Uppsala University, 751 05 Uppsala, Sweden}
\affiliation{Taras Shevchenko National University of Kyiv, Kiev, 01601, Ukraine}
\affiliation{Lancaster University, Lancaster LA1 4YB, United Kingdom}
\affiliation{Imperial College London, London SW7 2AZ, United Kingdom}
\affiliation{The University of Manchester, Manchester M13 9PL, United Kingdom}
\affiliation{University of Arizona, Tucson, Arizona 85721, USA}
\affiliation{University of California Riverside, Riverside, California 92521, USA}
\affiliation{Florida State University, Tallahassee, Florida 32306, USA}
\affiliation{Fermi National Accelerator Laboratory, Batavia, Illinois 60510, USA}
\affiliation{University of Illinois at Chicago, Chicago, Illinois 60607, USA}
\affiliation{Northern Illinois University, DeKalb, Illinois 60115, USA}
\affiliation{Northwestern University, Evanston, Illinois 60208, USA}
\affiliation{Indiana University, Bloomington, Indiana 47405, USA}
\affiliation{Purdue University Calumet, Hammond, Indiana 46323, USA}
\affiliation{University of Notre Dame, Notre Dame, Indiana 46556, USA}
\affiliation{Iowa State University, Ames, Iowa 50011, USA}
\affiliation{University of Kansas, Lawrence, Kansas 66045, USA}
\affiliation{Louisiana Tech University, Ruston, Louisiana 71272, USA}
\affiliation{Northeastern University, Boston, Massachusetts 02115, USA}
\affiliation{University of Michigan, Ann Arbor, Michigan 48109, USA}
\affiliation{Michigan State University, East Lansing, Michigan 48824, USA}
\affiliation{University of Mississippi, University, Mississippi 38677, USA}
\affiliation{University of Nebraska, Lincoln, Nebraska 68588, USA}
\affiliation{Rutgers University, Piscataway, New Jersey 08855, USA}
\affiliation{Princeton University, Princeton, New Jersey 08544, USA}
\affiliation{State University of New York, Buffalo, New York 14260, USA}
\affiliation{University of Rochester, Rochester, New York 14627, USA}
\affiliation{State University of New York, Stony Brook, New York 11794, USA}
\affiliation{Brookhaven National Laboratory, Upton, New York 11973, USA}
\affiliation{Langston University, Langston, Oklahoma 73050, USA}
\affiliation{University of Oklahoma, Norman, Oklahoma 73019, USA}
\affiliation{Oklahoma State University, Stillwater, Oklahoma 74078, USA}
\affiliation{Oregon State University, Corvallis, Oregon 97331, USA}
\affiliation{Brown University, Providence, Rhode Island 02912, USA}
\affiliation{University of Texas, Arlington, Texas 76019, USA}
\affiliation{Southern Methodist University, Dallas, Texas 75275, USA}
\affiliation{Rice University, Houston, Texas 77005, USA}
\affiliation{University of Virginia, Charlottesville, Virginia 22904, USA}
\affiliation{University of Washington, Seattle, Washington 98195, USA}
\author{V.M.~Abazov} \affiliation{Joint Institute for Nuclear Research, Dubna 141980, Russia}
\author{B.~Abbott} \affiliation{University of Oklahoma, Norman, Oklahoma 73019, USA}
\author{B.S.~Acharya} \affiliation{Tata Institute of Fundamental Research, Mumbai-400 005, India}
\author{M.~Adams} \affiliation{University of Illinois at Chicago, Chicago, Illinois 60607, USA}
\author{T.~Adams} \affiliation{Florida State University, Tallahassee, Florida 32306, USA}
\author{J.P.~Agnew} \affiliation{The University of Manchester, Manchester M13 9PL, United Kingdom}
\author{G.D.~Alexeev} \affiliation{Joint Institute for Nuclear Research, Dubna 141980, Russia}
\author{G.~Alkhazov} \affiliation{Petersburg Nuclear Physics Institute, St. Petersburg 188300, Russia}
\author{A.~Alton$^{a}$} \affiliation{University of Michigan, Ann Arbor, Michigan 48109, USA}
\author{A.~Askew} \affiliation{Florida State University, Tallahassee, Florida 32306, USA}
% ...
\collaboration{The D0 Collaboration\footnote{with visitors from
$^{a}$Augustana College, Sioux Falls, SD 57197, USA,
$^{b}$The University of Liverpool, Liverpool L69 3BX, UK,
$^{c}$Deutshes Elektronen-Synchrotron (DESY), Notkestrasse 85, Germany,
$^{d}$CONACyT, M-03940 Mexico City, Mexico,
$^{e}$SLAC, Menlo Park, CA 94025, USA,
$^{f}$University College London, London WC1E 6BT, UK,
$^{g}$Centro de Investigacion en Computacion - IPN, CP 07738 Mexico City, Mexico,
$^{h}$Universidade Estadual Paulista, S\~ao Paulo, SP 01140, Brazil,
$^{i}$Karlsruher Institut f\"ur Technologie (KIT) - Steinbuch Centre for Computing (SCC),
D-76128 Karlsruhe, Germany,
$^{j}$Office of Science, U.S. Department of Energy, Washington, D.C. 20585, USA,
$^{k}$American Association for the Advancement of Science, Washington, D.C. 20005, USA,
$^{l}$Kiev Institute for Nuclear Research (KINR), Kyiv 03680, Ukraine,
$^{m}$University of Maryland, College Park, MD 20742, USA,
$^{n}$European Orgnaization for Nuclear Research (CERN), CH-1211 Geneva, Switzerland,
$^{o}$Purdue University, West Lafayette, IN 47907, USA,
and
$^{p}$Institute of Physics, Belgrade, Belgrade, Serbia.
$^{\ddag}$Deceased.
}} \noaffiliation
\vskip 0.25cm

\date{March 20, 2017}

\hyphenation{un-cer-tain-ty}
           
\begin{abstract}
We present a combination of measurements of the top quark mass by the \dzero\ experiment in the lepton+jets and dilepton channels.
We use all the data collected in \RunI\ (1992--1996) at $\sqrts=1.8~\tev$ and  \RunII\ (2001--2011) at  $\sqrts=1.96~\tev$ of the 
Tevatron $\ppbar$ collider,
corresponding to integrated luminosities of 0.1~\invfb\ and 9.7~\invfb, respectively.
The  combined result is:  \mt= \result= \resulttot.

\end{abstract}


\maketitle

\section{Introduction}

The top quark is the heaviest known elementary particle with a mass approximately twice that of the electroweak vector bosons, and factor of 1.4 larger than that of the more recently discovered Higgs boson~\cite{Aad:2015zhl}.
Within the  standard model (SM),  this large mass arises from a large Yukawa coupling ($\approx 0.9$) to the Higgs field.  Consequently, loops involving the top quark contribute significantly to
electroweak quantum corrections, and therefore a precise measurement of the top quark mass, $\mt$, provides a means to test the
consistency of the SM.
Furthermore, the precise values of both the mass of the Higgs boson and the Yukawa coupling of the top quark may play a critical role in the history and stability of the universe (see e.g., Ref.~\cite{Degrassi:2012ry}).

The top quark was discovered in 1995 by the CDF and \dzero\ experiments during  \RunI\ (1992--1996) of the Fermilab Tevatron \ppbar\ collider at $\sqrts=1.8~\tev$~\cite{Abe:1995hr,Abachi:1995iq}.
\RunII\ (2001--2011) at $\sqrts=1.96~\tev$ followed, providing a factor of $\approx 150$ more top-antitop quark pairs than  \RunI, and far more precise measurements of $\mt$.
Using \ttbar\ events produced in the \dzero\ detector~\cite{run1det,run2det,Abolins2008,Angstadt2010},
we have measured $\mt$ in different decay channels~\cite{Mtop1-D0-di-l-PRL,Mtop1-D0-di-l-PRD,Mtop1-D0-l+jt-new1,Mtop2-D0-di-l-Nu-PLB,Mtop2-D0-di-l-ME-PRD,Mtop2-D0-l+jt-PRL,Mtop2-D0-l+jt-PRD}
using the full integrated luminosity of \RunI\ ($\int \mathcal{L}\;dt= 0.1~\invfb$) and \RunII\ ($\int \mathcal{L}\;dt=9.7~\invfb$). This article reports
the combination of these direct top quark mass measurements.

Direct measurements of the top quark mass have also been performed by the CDF experiment (see \eg~Ref.~\cite{cdf_latest_top_mass}) at the Tevatron, and by the ATLAS (see \eg~Ref.~\cite{atlas_latest_top_mass}) and CMS (see \eg~Ref.~\cite{cms_latest_top_mass})  experiments at the CERN LHC. In 2012, the Tevatron experiments  combined their measurements in Ref.~\cite{TeVTopComboPRD}
with the result $\mt=173.18 \pm 0.94~\gevcs$.
In 2014, a preliminary combination of ATLAS, CDF, CMS, and \dzero\ measurements~\cite{worldcombo}  yielded $\mt=173.34 \pm 0.76~\gevcs$. Both combinations are by now outdated as they do not include the latest and more precise measurements,
in particular, the final \dzero~\RunII~measurements discussed in this article.

The top quark mass is a fundamental free parameter of the SM. However, its definition depends on the scheme of theoretical calculations used for the perturbative expansion in quantum
chromodynamics (QCD). The inputs to the combination presented in this article are the direct measurements calibrated using Monte Carlo (MC) simulations. Hence, the measured mass corresponds to the MC mass parameter.
However, because of the presence of long range effects in QCD,
the relationship between the MC mass and other mass definitions, such as the pole mass or the mass in the  modified minimal subtraction ($\overline{\rm{MS}}$) scheme,
is not well established and has been  subject to debate for many years (see e.g., Ref.~\cite{Juste:2013dsa} and references therein).  
A recent work obtains a difference of +0.6~GeV between the MC mass and the pole mass in the context of an $e^+e^- \to \ttbar$  simulation with an uncertainty of 0.3~GeV~\cite{Butenschoen:2016lpz}.
Further studies are needed to produce a similar estimate  in the  context of  $\ppbar\to\ttbar$ production.

In Ref.~\cite{massxs}, we extracted
 the pole mass of the top quark from the measured $t\bar t$ cross section~\cite{mass-from-xsec-theory}.
  However, due to the ambiguity between the MC and pole mass,
the difficulty of properly assessing correlations between systematic uncertainties,
and the  large uncertainty of the pole mass measurement,
the latter
is not part of the combination presented in this article.

This article is structured as follows: we  first  summarize the input measurements;
we subsequently  present the combination of \RunII\ dilepton measurements, which provides one of the inputs to the \dzero\ combination;
we then discuss the different uncertainty categories and their correlations, and conclude with the final combined result.

\section{Decay channels and input measurements}

\begingroup
\squeezetable
\begin{table*}[!htbp]
\caption[Input measurements]{Summary of the input measurements to the combination. We indicate the method used to extract the mass of the top quark from the data (see the corresponding references for further details).
}\label{tab:input_publication}
\newcolumntype{M}[1]{>{\centering}m{#1}}
\begin{ruledtabular}
\begin{tabular}{lccM{5cm}cc}
Period & Channel          & { $\int \mathcal{L}\;dt$ (\invfb)}& Method &   \mt (\gevcs)&    Reference\\ \hline 
\RunI& \dil      &   $0.1$  &    { Combination of matrix weighting and neutrino weighting } & $168.4\phantom{0} \pm12.3\phantom{0}{\rm\ (stat)}\pm 3.6\phantom{0}{\rm\ (syst)}$  &\cite{Mtop1-D0-di-l-PRL,Mtop1-D0-di-l-PRD} \\
\RunI& \lplus    &   $0.1$  &    Matrix element & $180.1\phantom{0} \pm\phantom{0}3.6\phantom{0}{\rm\ (stat)}  \pm 3.9\phantom{0}{\rm\ (syst)}$ &\cite{Mtop1-D0-l+jt-new1} \\
\RunII& \dil      &   $9.7$  &    Neutrino weighting  &  $173.32\pm \phantom{0}1.36{\rm\ (stat)}\pm 0.85{\rm\ (syst)}$  &\cite{Mtop2-D0-di-l-Nu-PLB} \\
\RunII& \dil      &   $9.7$  &    Matrix element &  $173.93\pm \phantom{0}1.61{\rm\ (stat)}\pm 0.88{\rm\ (syst)}$&\cite{Mtop2-D0-di-l-ME-PRD} \\

\RunII& \lplus    &   $9.7$  &    Matrix element &   $174.98\pm \phantom{0}0.41{\rm\ (stat)}\pm 0.63{\rm\ (syst)}$ &\cite{Mtop2-D0-l+jt-PRL,Mtop2-D0-l+jt-PRD} \\
\end{tabular}
\end{ruledtabular}
\end{table*}
\endgroup
\label{sec:inputs}

To measure the top quark mass, we use $p\bar p \to t\bar t$ events
and assume that the top and antitop quark masses are equal~\cite{Abazov:2011ch,Aaltonen:2012zb,Aad:2013eva,Chatrchyan:2016mqq}.
Within the SM, the top quark decays into a  $W$ boson and a $b$ quark
almost 100\% of the time. Different channels arise from the possible decays of the pair of $W$ bosons:
\begin{enumerate}\renewcommand{\theenumi}{\roman{enumi}}

\item The ``dilepton'' channel (\dil)
corresponds to events  ($\approx 4.5\%$ of the total) where both $W$ bosons decay  into electrons or muons. 
This channel is quite free from background
but has a small yield.
The background is mainly due  to  $Z$+jets production,  but also receives contributions from  diboson ($WW$, $WZ$, $ZZ$), $W$+jets, and multijet production.

\item  The ``lepton+jets'' channel (\lplus) 
  corresponds to events ($\approx 30\%$ of the total) where one $W$ boson decays into $q\bar {q^{\prime}}$ and the other into an electron or a muon and a neutrino.
  This channel has a moderate yield and a background arising from
  $W$+jets production, $Z$+jets production, and multijet processes. 

\item
The ``all jets'' channel ($\approx 46\%$ of the total)
has events in which both $W$ bosons decay to $q\bar {q^{\prime}}$ that evolve into jets.
The yield is high, but the background  from multijet production is very large.

\item  
The ``tau channel'' ($\approx 20\%$ of the total) arises from events in which at least one of the $W$ bosons decays into $\tau\nu_{\tau}$. As the decays  $\tau\to\rm{hadrons}+\nu_{\tau}$  are difficult to distinguish from QCD jets,
 it is  not exploited for the top quark mass measurement.
However, the  $\tau\to\ell\nu_{\ell}\nu_{\tau}$ decays provide
contributions to the \dil\ and \lplus\ channels.

\end{enumerate}

The high mass of the top quark means that
the decay products tend to have high transverse momenta ($\pt$) relative to the beam axis  and large angular separations.
Reconstructing and identifying $t\bar t$ events requires
reconstruction and identification  of high $\pt$ electrons,
muons, and jets, and the measurement of the imbalance in transverse momentum in each event ($\met$) due to escaping neutrinos. 
In addition, identifying  $b$ jets is  an effective way of improving the purity of the selections.
Good momentum  resolution  is  required  for all these objects, and
the jet energy scale (JES)  has to be known with high precision.
In the  \RunII\ \lplus\ measurements, the uncertainty in the JES is reduced by performing an {in situ} calibration, which
exploits the $W\to q\bar{ q^{\prime}}$ decay by requiring the mass of the corresponding dijet system to be consistent with the mass of the $W$ boson ($ 80.4~\gevcs$).
This calibration, determined using light-quark jets (including charm jets), is  applied to jets of all flavors associated with $\ttbar$ decay.
It is then propagated to the
\RunII\ \dil\ measurements.

The  input measurements of $\mt$ for the presented combination are shown
in Table~\ref{tab:input_publication},
and consist of measurements performed during \RunI\ and \RunII\ in the
 \dil\   and \lplus\ channels using the full data sets.
\dzero\ also measured the top quark mass using the ``all jets'' channel in \RunI~\cite{Mtop1-D0-allh-PRL};
however, this measurement is not considered in the combination because  its uncertainty  is large
and  some subcomponents of the systematic uncertainty are not available.
Just as in \RunI,
two \dil\  mass measurements were performed in \RunII\ using
a neutrino weighting~\cite{Mtop2-D0-di-l-Nu-PLB}
technique (NW) and a matrix element method (ME)~\cite{Mtop2-D0-di-l-ME-PRD}. We discuss their combination
in the following section.

To combine the $\mt$ measurements, we use the Best Linear Unbiased Estimate (BLUE)~\cite{Valassi:2003mu}, assuming Gaussian  uncertainties,
both for the \dil\ \RunII\ and the final \dzero\ combinations.

\section{Combination of \RunII\ dilepton measurements}

In the \dil\ channel, the presence of two undetected neutrinos with  high \pt\ 
makes it impossible to fully reconstruct the kinematics of the final state.
To overcome this problem,
we use two  methods in \RunII.
The NW measurement~\cite{Mtop2-D0-di-l-Nu-PLB} is based on 
a weight function for each event which is computed
by comparing  the $x$-- and $y$-- components of the observed \met\ and
the hypothesized \pt\ components of the neutrinos, integrating over the  neutrino pseudorapidities~\cite{Note0}.
The maximum weight value 
indicates the most likely value of \mt\ in that event.
The first and second moments of this function are retained as the event-by-event variables sensitive to $\mt$.
Their distributions in MC events are used to form two-dimensional templates that depend upon the value of $\mt$. The templates are compared to the data to extract $\mt$. The ME~\cite{Mtop2-D0-di-l-ME-PRD} measurement  uses per-event probability densities, 
based on the reconstructed kinematic information,  obtained by 
integrating over the differential cross sections for the processes contributing to the observed events,
using leading order matrix elements for the $\ttbar$ production process and
accounting for detector resolution.
The unmeasured  neutrino momentum components are integrated out in this computation. The probability densities from all data events are combined to form a likelihood as a function of $\mt$, which is then maximized to determine $\mt$.

\subsection{Statistical uncertainties and correlation}
\label{sec:dilepton_correlation}

\begin{figure}[!ht]
  \centering
    \includegraphics[width=0.9\columnwidth]{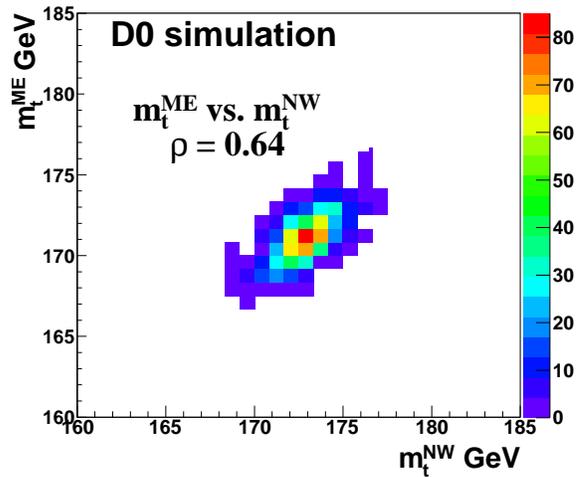}
\caption{
Two-dimensional distribution in the top quark masses
extracted from the MC event ensembles in the ME and NW analyses. The statistical correlation  $\rho$  is obtained from this distribution.}
\label{fig:2dcorr}
\end{figure}

The statistical uncertainties of the individual NW  and  ME  measurements are given in Table~\ref{tab:dilepton}.
Both  measurements are carried out using 
the same full \dzero\ \RunII\ data set, and similar selection criteria. Approximately 90\% of the selected events are common to both analyses, and the measurements
are therefore statistically correlated.
We use an ensemble testing method to estimate these  correlations.
In the first step, we generate 1000 ensembles of simulated background and signal  events with mass \mt=172.5 GeV that pass the criteria
of either the NW or the ME selection (see Refs.~\cite{Mtop2-D0-di-l-Nu-PLB} and ~\cite{Mtop2-D0-di-l-ME-PRD} for the detailed descriptions of the selections).
Each ensemble is generated with the same  number of events as observed in data,
using the expected signal and background fractions, separately for the \ee, \mumu, and \emu\ channels.  The ME and NW ensembles are then obtained using the individual and slightly more restrictive selection criteria from each analysis, and $\mt$ is extracted following each of the analysis methods.
From the two-dimensional distribution of the measured masses shown in Fig~\ref{fig:2dcorr},
we obtain a statistical correlation of $\rho=0.64 \pm 0.02$ between the two sets of measurements.

\subsection{Systematic uncertainties in \dil\ channel}

The different contributions to the systematic uncertainty  considered in the NW and ME measurements are reported in Table~\ref{tab:dilepton}. The sources of uncertainty are listed
in the following and briefly described when the naming is not self-explanatory.
More detailed descriptions are given in Refs.~\cite{Mtop2-D0-di-l-Nu-PLB} and ~\cite{Mtop2-D0-di-l-ME-PRD}, and in Sec.~\ref{sec:uncertainty} for the signal modeling uncertainties.

\begin{description}
\item[In situ light-jet calibration:] The statistical uncertainty of the JES calibration,  determined in the \lplus\  measurement using light-quark jets, and propagated to the \dil\ measurements.
\item[Response to \boldmath{$b$}, \boldmath{$q$}, and \boldmath{$g$} jets:] The part of the JES uncertainty that originates from
    differences in detector response among $b$, light-quark, and gluon jets.\item[Model for  \boldmath{$b$} jets:] The part of the JES uncertainty that originates from
    uncertainties specific to the modeling of $b$ jets.  This includes the dependence on semileptonic branching fractions and 
    modeling of $b$ quark fragmentation.
    
\item[Light-jet response:] The part of the JES
    uncertainty that affects all jets and includes the dependence of the calibration upon jet energy and pseudorapidity, and the effect of the out-of-cone calorimeter showering correction.

  \item[Jet energy resolution]\item[Jet identification efficiency]
\item[Multiple interaction model:]  The  systematic uncertainty that arises from modeling 
  the distribution of the number of interactions per Tevatron bunch crossing. 

\item[\boldmath{$b$} tag modeling:]

  The uncertainty related to the modeling of the $b$ tagging  efficiency
   for $b$, $c$, and light-flavor jets in MC simulation relative to data.
  
\item[Electron energy resolution]\item[Muon  momentum resolution]\item[Lepton momentum scale:]  The uncertainty arising from the calibration of electron  energy and muon momentum scales.

\item[Trigger efficiency:]
  The uncertainties in the estimation of lepton-based trigger efficiencies.

\item[Higher-order corrections:]  The modeling of  higher-order corrections in the simulation of \ttbar\ samples,  obtained from the difference between the next-to-leading-order \textsc{MC@NLO}~\cite{MCNLO}   and  the leading-order \textsc{ALPGEN}~\cite{ALPGEN} event generators.

\item[Initial and final state radiation:] The uncertainty due to the modeling of initial  and final state gluon radiation.
  
\item[Hadronization and underlying events:] The uncertainty associated with the modeling of hadronization and the underlying event, estimated from the difference between
different hadronization models.
\item[Color reconnection:]  The uncertainty due to the model of color reconnection.
\item[PDF:] The uncertainty from  the choice of parton density functions.
\item[Transverse momentum of \ttbar\ system:]  The uncertainty in the modeling of the distribution of the \pt\ of the \ttbar\ system. 

 \item[Yield of vector boson + heavy flavor:] The uncertainty associated with the production cross section for $Z$+$b\bar b$ and $Z$+$ c\bar c$ relative to $Z$+jets events.

\item[Background from simulation:] 
The systematic uncertainty on the  MC background, which  includes the uncertainty  from detector effects and the theoretical cross section. It does not include the uncertainties on the ratios of  $Z$+$b\bar b$ and $Z$+$ c\bar c$  to $Z$+jets cross sections, which belong to the previous category.

\item[Background based on data:] 
The  uncertainties from the modeling of the
  multijet and $W$+jets backgrounds estimated using data.

\item[Template statistics:]
  In the NW measurement, this uncertainty arises from the statistical fluctuations of individual bins in signal and background templates. In the  ME measurement, there is no such uncertainty as there is no template used to fit  the data.

       \item[Calibration method:]
  The calibration for both ME and NW measurements is determined
  using an ensemble  testing method. We generate  pseudo-experiments
  with the same number of events  as observed in data,  using MC events
  for signal and both MC and data-based samples  for backgrounds.
  Ensembles at different top quark mass hypotheses are generated  to
  determine a linear relation between the uncorrected measurement and
  the actual MC  mass, \ie, to determine  slope and offset parameters.
  The uncertainty in the  calibration method arises from the  uncertainty in
  the slope and offset parameters due  to the limited size of the MC and
  data-based samples.

\end{description}

All systematic uncertainties are considered as fully correlated between ME and NW except for the calibration method uncertainty,
as  the calibrations were performed using almost independent event samples.

The differences between the ME and NW  uncertainties reported in  Table~\ref{tab:dilepton} are consistent with
the expected statistical fluctuations in the various estimates.
The fluctuations are $\approx $~0.05--0.10~$\gevcs$, depending on the source, and their overall contributions  are well below the total uncertainties.  They therefore have a negligible impact on the overall uncertainties in
the individual measurements and  their combination.

\label{sec:dilepton_syst}
\begingroup
\squeezetable
\begin{table}[!h!tbp]
\caption[Input measurements]{Measurements in the \dil\ channel with contributions to the uncertainties, and  their combination.
The total systematic uncertainty   and the total uncertainty are obtained by adding the relevant contributions 
  in quadrature. All values are given in $\gevcs$. The symbol ``n/a'' stands for ``not applicable''.}\label{tab:dilepton}
\begin{center}
\renewcommand{\arraystretch}{1.30}
  \begin{tabular}{lccc} 
\hline \hline
       & \RunII &\RunII &   \RunII \\
       &   ME &    NW &   $\dil$ combination \\
\hline
\makeatletter{}top quark mass                                       &     173.93 &     173.32 &     173.50 \\ 
\hline
{In situ} light-jet calibration                  &       \phantom{00}0.46 &       \phantom{00}0.47 &       \phantom{00}0.47 \\ 
Response to $b$, $q$, and $g$ jets                     &       \phantom{00}0.30 &       \phantom{00}0.27 &       \phantom{00}0.28 \\ 
Model for $b$ jets                               &       \phantom{00}0.21 &       \phantom{00}0.10 &       \phantom{00}0.13 \\ 
Light-jet response                               &       \phantom{00}0.20 &       \phantom{00}0.36 &       \phantom{00}0.31 \\ 
Jet energy resolution                            &       \phantom{00}0.15 &       \phantom{00}0.12 &       \phantom{00}0.13 \\ 
Jet identification efficiency                    &       \phantom{00}0.08 &       \phantom{00}0.03 &       \phantom{00}0.04 \\ 
Multiple interaction   model                     &       \phantom{00}0.10 &       \phantom{00}0.06 &       \phantom{00}0.07 \\ 
$b$ tag modeling                                 &       \phantom{00}0.28 &       \phantom{00}0.19 &       \phantom{00}0.22 \\ 
Electron energy resolution                       &       \phantom{00}0.16 &       \phantom{00}0.01 &       \phantom{00}0.05 \\ 
Muon momentum   resolution                       &       \phantom{00}0.10 &       \phantom{00}0.03 &       \phantom{00}0.05 \\ 
Lepton momentum scale                            &       \phantom{00}0.10 &       \phantom{00}0.01 &       \phantom{00}0.04 \\ 
Trigger efficiency                               &       \phantom{00}0.06 &       \phantom{00}0.06 &       \phantom{00}0.06 \\ 
Higher-order corrections                         &       \phantom{00}0.16 &       \phantom{00}0.33 &       \phantom{00}0.28 \\ 
Initial and final state radiation                &       \phantom{00}0.16 &       \phantom{00}0.15 &       \phantom{00}0.15 \\ 
Hadronization and underlying event               &       \phantom{00}0.31 &       \phantom{00}0.11 &       \phantom{00}0.17 \\ 
 Color reconnection                              &       \phantom{00}0.15 &       \phantom{00}0.22 &       \phantom{00}0.20 \\ 
PDF                  &       \phantom{00}0.20 &       \phantom{00}0.08 &       \phantom{00}0.11 \\ 
Transverse momentum of $\ttbar$ system               &       \phantom{00}0.03 &       \phantom{00}0.07 &       \phantom{00}0.06 \\ 
Yield of vector boson + heavy flavor                  &       \phantom{00}0.06 &       \phantom{00}0.04 &       \phantom{00}0.05 \\ 
Background from simulation                       &       \phantom{00}0.06 &       \phantom{00}0.01 &       \phantom{00}0.02 \\ 
Background based on data                           &       \phantom{00}0.07 &       \phantom{00}0.00 &       \phantom{00}0.02 \\ 
Template statistics                   & \phantom{00}n/a & \phantom{00}0.18 & \phantom{00}0.13 \\
Calibration method                               &       \phantom{00}0.03 &       \phantom{00}0.07 &       \phantom{00}0.05 \\ 
\hline
Systematic uncertainty                             &       \phantom{00}0.88 &       \phantom{00}0.85 &       \phantom{00}0.84 \\ 
Statistical uncertainty                          &       \phantom{00}1.61 &       \phantom{00}1.36 &       \phantom{00}1.31 \\ 
\hline
Total uncertainty                                &       \phantom{00}1.84 &       \phantom{00}1.61 &       \phantom{00}1.56 \\ 
 
\hline\hline
\end{tabular}
\end{center}
\end{table}
\endgroup
\subsection{\dil\ combination}
To obtain the ME and NW combination through the BLUE method
we use the correlations and uncertainties discussed in 
 Sec.~\ref{sec:dilepton_correlation} and Sec.~\ref{sec:dilepton_syst}.

The result of the BLUE combination is  $\mt=173.50\pm 1.31\,{\rm\ (stat)}\pm 0.84{\,\rm\ (syst)}~\gevcs$.
The breakdown of uncertainties is given in Table~\ref{tab:dilepton}.
The  weights for the NW and ME measurements are 71\% and 29\%, respectively.
The NW and ME measurements agree with a $\chi^2$ of   0.2 for one degree of freedom, corresponding to a probability of 65\%.
As a test of stability, we  change the statistical correlation between the two methods from  0.50 to 0.70
to conservatively cover the range of systematic and statistical uncertainty in its determination.
The resulting \mt\  changes by less than 0.04~\gevcs.

This combination of the \RunII\ \dil\ measurements is used as an input to the overall  combination discussed in the next sections.

\section{Uncertainty categories in the overall combination}
\label{sec:uncertainty}

For the overall combination, the systematic uncertainties are grouped into sources of same or similar origin to form  uncertainty categories. We employ categories similar to
those used in the  Tevatron
top quark mass combination~\cite{TeVTopComboPRD}  and use the same naming scheme.

\begin{description}
  \item[{In situ} light-jet calibration:] The part of the
   JES uncertainty that originates from the
   {in situ} calibration procedure using light-quark jets. This uncertainty has a statistical origin.
            For the \RunII\ $\dil$ measurement, the uncertainty 
   from transferring the $\lplus$ calibration to the dilepton event topology
   is included in the {light-jet response } category described below.
 
\item[Response to \boldmath{$b$}, \boldmath{$q$}, and \boldmath{$g$} jets:]
 As described in Sec.~\ref{sec:dilepton_syst}.
       \item[Model for \boldmath{$b$} jets:]
 As described in Sec.~\ref{sec:dilepton_syst}.

  \item[Light-jet response:] The part of the JES
    uncertainty that includes  calibrations of the
    absolute  energy-dependent response and the relative $\eta$-dependent response,
    and, for \RunII, the out-of-cone     calorimeter showering correction. This uncertainty applies to  jets of any flavor.

  \item[Out-of-cone correction:] The part of the JES uncertainty that originates from
    modeling of  uncertainties associated with light-quark 
    fragmentation and out-of-cone calorimeter showering corrections in \RunI\ measurements. For \RunII\ measurements,
    it is included in the { light-jet response} category.

  \item[Offset:] This     includes the uncertainty
    arising from uranium noise in the \dzero\ calorimeter and from the
       corrections to the JES due to multiple interactions.     While such uncertainties were sizable in \RunI,
    the shorter integration time in the
    calorimeter electronics  and the {in situ} JES calibration make them negligible  in     \RunII.
 
  \item[Jet modeling:] 
The systematic uncertainties arising from uncertainties  in jet resolution
and identification. 
  \item[Multiple interactions model:]
 As described in Sec.~\ref{sec:dilepton_syst}.

\item[\boldmath{$b$} tag modeling:]
    As described in Sec.~\ref{sec:dilepton_syst}. 

  \item[Lepton modeling:] The uncertainties in the modeling
    of the scale and resolution of lepton \pt, which were 
taken to be negligible
in  \RunI.   \item[Signal modeling:] The systematic uncertainties arising from
     \ttbar\ event modeling, which are correlated across all
    measurements. 
This includes the sources described below.
In Run~I, the breakdown into the first four items could not be performed,
because the MC generators used at that time did not have the same flexibility as the more modern generators. Instead,
the overall signal modeling uncertainty was estimated by  changing the main parameters of a MC generator or comparing results from two different generators.
  \begin{enumerate}\renewcommand{\theenumi}{\roman{enumi}}

  \item The uncertainty associated with the modeling of initial and final state radiation,
        obtained by changing
      the renormalization scale in the       scale-setting procedure  relative to its default, as suggested in Ref.~\cite{Cooper:2011gk}. Studies of $Z\to\ell\ell$ data indicate
                that a range of variation between factors of $\frac 1 2$ and 2
      of this scale covers the  mismodeling~\cite{Mtop2-D0-l+jt-PRD}.

    \item The uncertainty 
    from higher-order corrections evaluated from a comparison of \ttbar\ samples generated using {\textsc MC@NLO}~\cite{MCNLO} and 
    {\textsc ALPGEN}~\cite{ALPGEN}, both interfaced to {\textsc HERWIG}~\cite{HERWIG5,HERWIG6} for the simulation of parton showers and hadronization.

\item    The systematic uncertainty arising from a change in the 
  phenomenological description of color reconnection (CR) among final state  partons~\cite{CR}.   It is obtained from the difference between event samples generated using {\textsc PYTHIA}~\cite{pythia}  with the Perugia 2011 tune and
    using {\textsc PYTHIA} with the  Perugia 2011NOCR tune~\cite{Skands2010}.
\item The systematic 
uncertainty associated with the choice for modeling
parton-shower, hadronization, and underlying event.  It includes the changes observed when 
    substituting {\textsc PYTHIA}    for {\textsc HERWIG}~\cite{HERWIG5,HERWIG6} when 
    modeling \ttbar\ signal.

  \item The uncertainty associated with
  the choice of PDF used
  to generate the \ttbar\ MC events.
  It is estimated in \RunII\
 by changing the 20 eigenvalues of the {\textsc{CTEQ6.1M}} PDF~\cite{Nadolsky:2008zw}  within their uncertainties. In \RunI, it was obtained by comparing 
 {\textsc{CTEQ3M}}~\cite{Lai:1994bb} with {\textsc{MRSA}}~\cite{Martin:1994kn} for \dil, and {\textsc{CTEQ4M}}~\cite{Lai:1996mg} with {\textsc{CTEQ5L}}~\cite{Lai:1999wy} for \lplus\ events.

\end{enumerate}

  \item[Background from theory:] 
This systematic uncertainty on  background originating from theory 
takes into account the    
 uncertainty in modeling the background sources. It is correlated among
    all measurements in the same channel, and includes uncertainties on  background composition, normalization, and  distributions.

 \item[Background based on data:] This includes     uncertainties associated with the modeling of
    multijet background in the
 \lplus\ channel, and
       multijet and $W$+jets backgrounds in the \dil\ channel, which are estimated  using data. This also includes the effects of trigger uncertainties determined from the data.

  \item[Calibration method:] The uncertainty arising from any source specific
    to a particular fitting method, includes effects such as
      the finite number of MC events
    available to calibrate each method.

\end{description}

Table~\ref{tab:inputs} summarizes the 
input measurements
and their corresponding statistical and systematic uncertainties.

\section{Correlations}
\label{sec:corltns}

The following correlations are used to combine the measurements:
\begin{enumerate}\renewcommand{\theenumi}{\roman{enumi}}
  \item The uncertainties listed as `{statistical uncertainty}', 
    `{calibration method}', and `{background based on data}'
    are taken to be uncorrelated among the measurements.
  \item The uncertainties in the `{in situ} {light-jet  
    calibration}'
    category are taken to be correlated among the \RunII\ measurements
           since the
    $\dil$ measurement uses the JES calibration  determined in
   the $\lplus$ channel.  
  \item The uncertainties in  `{response to $b$, $q$, and $g$ jets}', `{jet modeling}',  `{$b$ tag modeling}',  `{multiple interaction model}', and `{lepton modeling}'
    are taken
    to be 100\% correlated among  \RunII\ measurements.

  \item The uncertainties in     `{out-of-cone correction}' and `{offset}'
categories  are taken to be 100\% correlated among  \RunI\ measurements.
    
  \item The uncertainties in     `{model for $b$ jets}' and `{signal modeling}'
categories  are taken to be 100\% correlated among all measurements.

  \item The uncertainties in `{light-jet response}'  are taken
    to be 100\% correlated among  the \RunI\ and  the \RunII\ measurements, but uncorrelated between \RunI\ and \RunII.

  \item The uncertainties in `{background  from theory}' are taken to be
    100\% correlated among all measurements in the same channel.

\end{enumerate}

A summary of the correlations among the different systematic categories is shown in Table~\ref{tab:correl}.
Using the inputs from Table~\ref{tab:inputs} and the correlations specified in Table~\ref{tab:correl},
we obtain an overall  matrix  of correlation coefficients in
Table~\ref{tab:coeff}.

\begingroup
\squeezetable
\begin{table}[!h!tbp]
\caption[Input measurements]{Summary of  measurements used to determine the
  \dzero\ average \mt.  Integrated luminosity ($\int \mathcal{L}\;dt$) has units of
  \invfb, and all other values  are in $\gevcs$.  The uncertainty categories and 
  their correlations are described in Sec.~\ref{sec:uncertainty}.  The total systematic uncertainty 
  and the total uncertainty are obtained by adding the relevant contributions 
  in quadrature. The symbol ``n/a'' stands for ``not applicable'', and the symbol ``n/e''  for ``not evaluated'' (but expected to be negligible).

}
\label{tab:inputs}
\begin{center}
\renewcommand{\arraystretch}{1.30}
\newcolumntype{H}{>{\setbox0=\hbox\bgroup}c<{\egroup}@{}}  \begin{tabular}{lccccH} 
\hline \hline
       & \multicolumn{2}{c}{\dzero\ {\RunI}} 
       & \multicolumn{2}{c}{\dzero\ {\RunII }}\\
   & $\lplus$ &  \multicolumn{1}{c}{$\dil$} & $\lplus$ & $\dil$  \\
\hline                                                                                                                                                       
$\int \mathcal{L}\;dt$ &    \phantom{00} 0.1\phantom{0}   &    \phantom{00} 0.1\phantom{0}    &    \phantom{00} 9.7\phantom{0}     &         \phantom{00} 9.7\phantom{0}       \\
\hline                                                                                               
\makeatletter{}top quark mass                                     & 180.10 & 168.40 & 174.98 & 173.50 & 174.95 \\ 
\hline
{In situ} light-jet calibration           & \phantom{00}n/a   & \phantom{00}n/a   & \phantom{00}0.41 & \phantom{00}0.47 & \phantom{00}0.41 \\ 
Response to $b$, $q$, and $g$ jets             & \phantom{00}n/e   & \phantom{00}n/e   & \phantom{00}0.16 & \phantom{00}0.28 & \phantom{00}0.16 \\ 
Model for $b$ jets                            &  \phantom{00}0.71 &  \phantom{00}0.71 &  \phantom{00}0.09 &  \phantom{00}0.13 &  \phantom{00}0.09 \\ 
Light-jet response                         &  \phantom{00}2.53 &  \phantom{00}1.12 &  \phantom{00}0.21 &  \phantom{00}0.31 &  \phantom{00}0.21 \\ 
Out-of-cone correction                        & \phantom{00}2.00 & \phantom{00}2.00 & \phantom{00}n/a   & \phantom{00}n/a   & $<0.01$ \\ 
Offset                                        & \phantom{00}1.30 & \phantom{00}1.30 & \phantom{00}n/a   & \phantom{00}n/a   & $<0.01$ \\ 
Jet modeling                                  & \phantom{00}n/e   & \phantom{00}n/e   & \phantom{00}0.07 & \phantom{00}0.14 & \phantom{00}0.07 \\ 
Multiple interaction  model                   & \phantom{00}n/e   & \phantom{00}n/e   & \phantom{00}0.06 & \phantom{00}0.07 & \phantom{00}0.06 \\ 
$b$ tag modeling                              & \phantom{00}n/e   & \phantom{00}n/e   & \phantom{00}0.10 & \phantom{00}0.22 & \phantom{00}0.10 \\ 
Lepton modeling                               & \phantom{00}n/e   & \phantom{00}n/e   & \phantom{00}0.01 & \phantom{00}0.08 & \phantom{00}0.01 \\ 
Signal modeling                               &  \phantom{00}1.10 &  \phantom{00}1.80 &  \phantom{00}0.35 &  \phantom{00}0.43 &  \phantom{00}0.35 \\ 
Background from theory                        &  \phantom{00}1.00 &  \phantom{00}1.10 &  \phantom{00}0.06 &  \phantom{00}0.05 &  \phantom{00}0.06 \\ 
Background based on data                      & \phantom{00}n/e   & \phantom{00}n/e   & \phantom{00}0.09 & \phantom{00}0.06 & \phantom{00}0.09 \\ 
Calibration method                            &  \phantom{00}0.58 &  \phantom{00}1.14 &  \phantom{00}0.07 &  \phantom{00}0.14 &  \phantom{00}0.07 \\ 
\hline
Systematic uncertainty                        &  \phantom{00}3.89 &  \phantom{00}3.63 &  \phantom{00}0.63 &  \phantom{00}0.84 &  \phantom{00}0.64 \\ 
Statistical uncertainty                       &  \phantom{00}3.60 &  \phantom{0}12.30 &  \phantom{00}0.41 &  \phantom{00}1.31 &  \phantom{00}0.40 \\ 
\hline
Total uncertainty                       & \phantom{00}5.30 &  \phantom{0}12.83 & \phantom{00}0.76 & \phantom{00}1.56 & \phantom{00}0.75 \\ 
 
\hline\hline
\end{tabular}
\end{center}
\end{table}
\begin{table}[h]
\begin{center}
\caption{
Summary of correlations among sources of uncertainty. The symbols $\times$ or $\otimes$ within any category 
indicate the uncertainties that are 100\% correlated.
The uncertainties marked as $\times$ are uncorrelated with those marked as $\otimes$.
The symbol $0$ indicates  absence of correlations.
The symbol ``n/a'' stands for ``not applicable''.
\label{tab:correl} }
\begin{tabular}{l c c c c}
\hline\hline
&\multicolumn{2}{c }{\dzero\ \RunI} & \multicolumn {2}{c}{\dzero\ \RunII} \\

      & \lplus & \dil &  \lplus & \dil \\
\hline
{In situ} light-jet calibration     	    & n/a  & n/a  & $\times$ & $\times$  \\
 response to $b$, $q$, and $g$ jets           &    n/a      &     n/a      & $\times$ & $\times$ \\ 
 Model for $b$ jets           & $\times$ & $\times$ & $\times$ & $\times$ \\ 
  Light-jet response              &    $\otimes$      &     $\otimes$     & $\times$ & $\times$ \\
 Out-of-cone correction             & $\times$ & $\times$ & n/a  & n/a  \\
Offset           & $\times$ & $\times$ & n/a & n/a\\
Jet modeling   &n/a  & n/a  & $\times$ & $\times$ \\
 Multiple interactions  model	    &   n/a       &     n/a   & $\times$ & $\times$ \\
$b$ tag modeling     &      n/a    &    n/a      & $\times$ & $\times$ \\
 Lepton modeling      &      n/a     &    n/a       & $\times$ & $\times$ \\
Signal modeling  & $\times$ & $\times$ & $\times$ & $\times$ \\
Background from theory   & $\times$ &     $\otimes$    & $\times$ &    $\otimes$     \\ 
Background based on data   &  n/a & n/a & 0 & 0\\
 Calibration method      &  0 & 0 & 0 & 0 \\
Statistical & 0 & 0 & 0 &0  \\ 
\hline
\end{tabular}
\end{center}
\end{table}
\endgroup

\begingroup
\squeezetable
\begin{table}[h]
\caption[Global correlations between input measurements]{The matrix of correlation coefficients used to determine the
  \dzero\ average top quark mass.}
\begin{center}
\renewcommand{\arraystretch}{1.30}
\label{tab:coeff}
\begin{tabular}{l|cccc}   
\makeatletter{} & \shortstack{\RunI,\\ $\lplus\phantom{^{\prime}}$ \,} & \shortstack{\RunI,\\  \dil \,} & \shortstack{\RunII,\\ $\lplus\phantom{^{\prime}}$ \,} & \shortstack{\RunII,\\   \dil \,} \\ 
\hline
\RunI, \lplus & 1.00 &       &       &       \\
\RunI,  \dil & 0.16 & 1.00 &       &       \\
\RunII, \lplus & 0.13 & 0.07 & 1.00 &       \\
\RunII,   \dil & 0.07 & 0.05 & 0.43 & 1.00 \\  
  
\end{tabular}
\end{center}

\end{table}
\endgroup

\section{Results}
\label{sec:results}

We combine the \dzero\ input measurements of Table~\ref{tab:inputs} using the BLUE method.
The BLUE combination has a $\chi^2$ of 2.5 for 3 degrees of freedom, corresponding to
a probability of 47\%.
The pulls and weights for each of the inputs obtained from the  BLUE method are listed in Table~\ref{tab:stat}.
Here, the pull associated to each input value $m_i$  with uncertainty $\sigma_i$ is calculated as $\frac {(m_i-\mt)}{\sqrt{\sigma_i^2-\sigma^2_{\mt}}}$, where $\sigma^2_{\mt}$
is the uncertainty in the combination, and indicates the degree of agreement of the input with the combined value.
The weight  $w_i$ given to the input measurement $m_i$ is $ w_i = \sum_{j=1}^4 (\mathrm{Cov}^{-1})_{ij} / N$, where   $\mathrm{Cov}$ is the covariance matrix of the input measurements, and $N$ is a normalization term ensuring
$\sum_{i=1}^4 w_i=1$.  The covariance matrix expressed in terms of the correlation coefficients between the measurements ${c}_{ij}$  (with the convention ${c}_{ii}=0$) is:
$\mathrm{Cov_{ij}}= {\sigma_i}{\sigma_j}( \delta_{ij} + {c}_{ij})$,  where $\delta_{ij}$ is the Kronecker $\delta$.
  At first order in the correlation coefficients,  its inverse is given by   $(\mathrm{Cov}^{-1})_{ij}= \frac{1}{\sigma_i}\frac{ 1}{\sigma_j}( \delta_{ij} - {c}_{ij})$, so that the weight $w_i$ can be written as  $w_i=\frac 1 {\sigma^2_i} ( 1 - \sum_{j\neq i}\frac {\sigma_i}{\sigma_j}{c}_{ij})/N'$, $N'$ being a normalization term.
 This expression shows that the weight for the \RunI\ \dil\  measurement is negative mainly because the correlation with the \RunII\ \lplus\  measurement  (0.07)  is  larger than the ratio  of their uncertainties  (0.76/12.7).

\begingroup
\squeezetable
\begin{table}[ht]
\caption[The pull and weight of each measurement]{The pull and weight for each
  input channel when using the BLUE method to
  determine the average top quark mass.
}
\begin{center}
\renewcommand{\arraystretch}{1.30}
  \begin{tabular}{lcccc} 
\hline \hline
       & \multicolumn{2}{c}{\dzero\ {\RunI}} 
       & \multicolumn{2}{c}{\dzero\ {\RunII }}\\
   & \lplus & \dil &  \lplus & \dil  \\
 \hline                                                       
\makeatletter{} 
Pull &   \phantom{$-$}0.98\phantom{0} & $-$0.51\phantom{0} &   \phantom{$-$}0.63\phantom{0} & $-$1.06\phantom{0} \\  
Weight\hspace{0.4cm} &  \phantom{$-$}0.002 & $-$0.003 &  \phantom{$-$}0.964 &  \phantom{$-$}0.035 \\
 
\hline    \hline    
\end{tabular}
\end{center}
\label{tab:stat} 
\end{table} 
\endgroup

The resulting combined value for the top quark mass is
\begin{eqnarray}
\nonumber
\mt=\result .
\end{eqnarray}
Adding the statistical and systematic uncertainties
in quadrature yields a total uncertainty of $\tot$~\gev, corresponding to a
relative precision of 0.43\% on the top quark mass. The breakdown of the uncertainties is 
shown in Table\,\ref{tab:BLUEuncert}. The dominant sources of uncertainty are the statistical uncertainty, the JES calibration, which has statistical origin, and the modeling of the signal.
 The total statistical and systematic  
uncertainties are reduced relative to  the published \dzero\ and CDF combination~\cite{TeVTopComboPRD} 
due primarily to the latest and most accurate \dzero\ \lplus\ analysis~\cite{Mtop2-D0-l+jt-PRL,Mtop2-D0-l+jt-PRD}.
As a test of stability, we vary the correlation of the dominant source of uncertainties, `signal modeling',  from 100\% to 0\%, first  between \RunI\ and \RunII\ measurements,
and  in a second check between all measurements.
The combined value of $\mt$ does not change by more than 50~MeV, while the uncertainty changes by no more than 20~MeV.
This is due to the fact that the \RunII\ \lplus\ measurement dominates the combination with a weight of 96\%. Thus, the combination is not sensitive to the detailed description of the correlation of systematic uncertainties.
Due to a much smaller total uncertainty resulting in the large weight for the \lplus\ measurement, the improvement in the combined uncertainty relative to the individual \lplus\ uncertainty is smaller than 10 MeV.

\begingroup
\squeezetable
\begin{table}[t]
\caption{\label{tab:BLUEuncert} 
Combination of \dzero\  measurements of \mt\ and contributions to its overall uncertainty.
The uncertainty categories are 
defined in the text. The total systematic uncertainty and the total
uncertainty are obtained  
by adding the relevant contributions in quadrature.}
\begin{center}
\renewcommand{\arraystretch}{1.30}
\newcolumntype{H}{>{\setbox0=\hbox\bgroup}c<{\egroup}@{}}  \begin{tabular}{lHHHHc} 
\hline
\hline
  &&&&& \dzero\ combined values (\gevcs) \\ \hline
\makeatletter{}top quark mass                                     & 180.10 & 168.40 & 174.98 & 173.50 & 174.95 \\ 
\hline
{In situ} light-jet calibration           & \phantom{00}n/a   & \phantom{00}n/a   & \phantom{00}0.41 & \phantom{00}0.47 & \phantom{00}0.41 \\ 
Response to $b$, $q$, and $g$ jets             & \phantom{00}n/e   & \phantom{00}n/e   & \phantom{00}0.16 & \phantom{00}0.28 & \phantom{00}0.16 \\ 
Model for $b$ jets                            &  \phantom{00}0.71 &  \phantom{00}0.71 &  \phantom{00}0.09 &  \phantom{00}0.13 &  \phantom{00}0.09 \\ 
Light-jet response                         &  \phantom{00}2.53 &  \phantom{00}1.12 &  \phantom{00}0.21 &  \phantom{00}0.31 &  \phantom{00}0.21 \\ 
Out-of-cone correction                        & \phantom{00}2.00 & \phantom{00}2.00 & \phantom{00}n/a   & \phantom{00}n/a   & $<0.01$ \\ 
Offset                                        & \phantom{00}1.30 & \phantom{00}1.30 & \phantom{00}n/a   & \phantom{00}n/a   & $<0.01$ \\ 
Jet modeling                                  & \phantom{00}n/e   & \phantom{00}n/e   & \phantom{00}0.07 & \phantom{00}0.14 & \phantom{00}0.07 \\ 
Multiple interaction  model                   & \phantom{00}n/e   & \phantom{00}n/e   & \phantom{00}0.06 & \phantom{00}0.07 & \phantom{00}0.06 \\ 
$b$ tag modeling                              & \phantom{00}n/e   & \phantom{00}n/e   & \phantom{00}0.10 & \phantom{00}0.22 & \phantom{00}0.10 \\ 
Lepton modeling                               & \phantom{00}n/e   & \phantom{00}n/e   & \phantom{00}0.01 & \phantom{00}0.08 & \phantom{00}0.01 \\ 
Signal modeling                               &  \phantom{00}1.10 &  \phantom{00}1.80 &  \phantom{00}0.35 &  \phantom{00}0.43 &  \phantom{00}0.35 \\ 
Background from theory                        &  \phantom{00}1.00 &  \phantom{00}1.10 &  \phantom{00}0.06 &  \phantom{00}0.05 &  \phantom{00}0.06 \\ 
Background based on data                      & \phantom{00}n/e   & \phantom{00}n/e   & \phantom{00}0.09 & \phantom{00}0.06 & \phantom{00}0.09 \\ 
Calibration method                            &  \phantom{00}0.58 &  \phantom{00}1.14 &  \phantom{00}0.07 &  \phantom{00}0.14 &  \phantom{00}0.07 \\ 
\hline
Systematic uncertainty                        &  \phantom{00}3.89 &  \phantom{00}3.63 &  \phantom{00}0.63 &  \phantom{00}0.84 &  \phantom{00}0.64 \\ 
Statistical uncertainty                       &  \phantom{00}3.60 &  \phantom{0}12.30 &  \phantom{00}0.41 &  \phantom{00}1.31 &  \phantom{00}0.40 \\ 
\hline
Total uncertainty                       & \phantom{00}5.30 &  \phantom{0}12.83 & \phantom{00}0.76 & \phantom{00}1.56 & \phantom{00}0.75 \\ 
 
\hline \hline
\end{tabular}        
\end{center}
\end{table}
\endgroup

\begin{figure*}[h]
\begin{center}
  \includegraphics[width=0.85\textwidth]{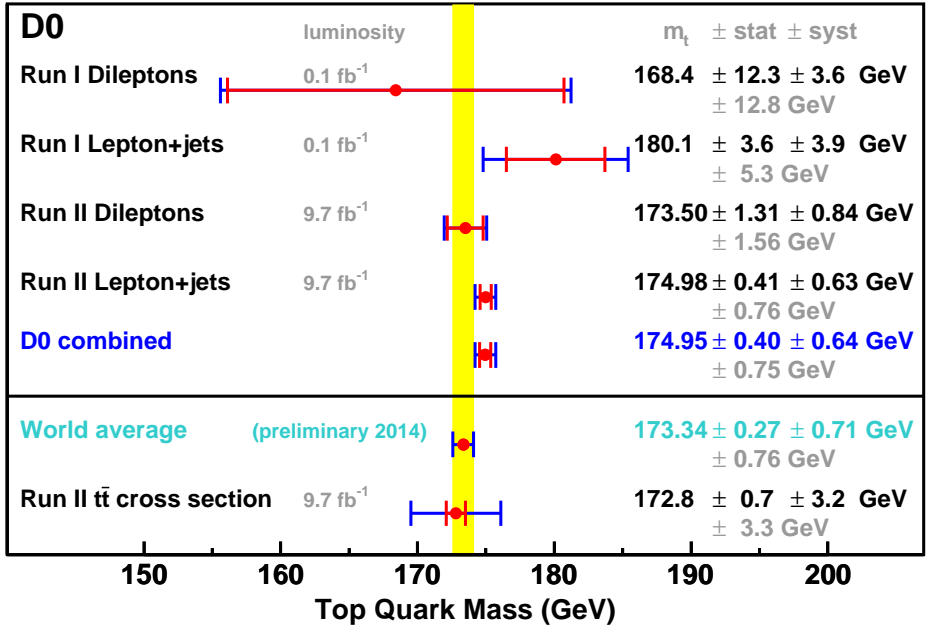}
  \end{center}
\vspace{-0.6cm}
    \caption{\label{fig:summary} A summary of the top quark mass measurements used 
    in the \dzero combination~\cite{Mtop1-D0-di-l-PRL,Mtop1-D0-di-l-PRD,Mtop1-D0-l+jt-new1,Mtop2-D0-di-l-Nu-PLB,Mtop2-D0-di-l-ME-PRD,Mtop2-D0-l+jt-PRL,Mtop2-D0-l+jt-PRD}, along with the \dzero final result,
    and the top quark pole mass extracted from the  \dzero\ cross section measurement~\cite{massxs}. The latter   is not used in the combination.
    The inner red uncertainty bars represent the statistical uncertainties, while the blue bars represent the  total uncertainties.
        For comparison, we also show the preliminary 2014 world average of \mt~\cite{worldcombo} which includes  \dzero\ \RunII\ \dil\ and \lplus\  measurements that are now superseded.
    For the top quark pole mass extracted from  the   \dzero\ cross section measurement, a 1.1~\gev\ theory uncertainty is included in the systematic uncertainty, and   the statistical uncertainty is determined such as  its relative contribution to the experimental uncertainty is the same as for the cross-section measurement. 
}
\end{figure*}

The input measurements and the resulting \dzero\ average mass of the top 
quark are summarized in Fig.~\ref{fig:summary}, along with  the top quark pole mass extracted by \dzero
from the measurement of the \ttbar\ cross section~\cite{massxs}.

\section{Summary}
\label{sec:summary}

We have presented the combination of the measurements of the top quark mass in all  \dzero\ data. 
 Taking into
account the statistical and systematic uncertainties and their
correlations, we find a combined average of
  $\mt=\resulttot$.
This measurement with, a relative precision of 0.43\%,
constitutes the legacy \RunI\ and \RunII\  measurement of the top quark mass in the \dzero\ experiment.

\section{Acknowledgments} \makeatletter{}
We thank the staffs at Fermilab and collaborating institutions,
and acknowledge support from the
Department of Energy and National Science Foundation (United States of America);
Alternative Energies and Atomic Energy Commission and
National Center for Scientific Research/National Institute of Nuclear and Particle Physics  (France);
Ministry of Education and Science of the Russian Federation, 
National Research Center ``Kurchatov Institute" of the Russian Federation, and 
Russian Foundation for Basic Research  (Russia);
National Council for the Development of Science and Technology and
Carlos Chagas Filho Foundation for the Support of Research in the State of Rio de Janeiro (Brazil);
Department of Atomic Energy and Department of Science and Technology (India);
Administrative Department of Science, Technology and Innovation (Colombia);
National Council of Science and Technology (Mexico);
National Research Foundation of Korea (Korea);
Foundation for Fundamental Research on Matter (The Netherlands);
Science and Technology Facilities Council and The Royal Society (United Kingdom);
Ministry of Education, Youth and Sports (Czech Republic);
Bundesministerium f\"{u}r Bildung und Forschung (Federal Ministry of Education and Research) and 
Deutsche Forschungsgemeinschaft (German Research Foundation) (Germany);
Science Foundation Ireland (Ireland);
Swedish Research Council (Sweden);
China Academy of Sciences and National Natural Science Foundation of China (China);
and
Ministry of Education and Science of Ukraine (Ukraine).
 
\bibliographystyle{apsrev_custom2}
\bibliography{references}

\end{document}